# Structure and spectroscopy characterization of $La_{1-x}Sm_xVO_4$ luminescent nanoparticles synthesized co-precipitation and sol-gel methods

O.V. Chukova[a], S.A. Nedilko[a], S.G. Nedilko[a], A. Papadopoulos[b], A.A. Slepets[a],

E.I. Stratakis[b], T.A. Voitenko[a]

[a] *Taras Shevchenko National University of Kyiv, Volodymyrska Str., 64/13, Kyiv 01601, Ukraine.*
[b] *Institute of Electronic Structure & Laser (IESL) of Foundation for Research & Technology Hellas (FORTH), Heraklion 71110, Crete, Greece*

**Abstract**

The Sm-activated orthvanadate nanoparticles with $La_{1-x}Sm_xVO_4$ (x ≤ 0.3) composition were synthesized by co-precipitation and sol-gel methods. XRD study has shown that synthesized samples are characterized by monoclinic or tetragonal structure as well as their mixture dependently on Sm concentration and methods of synthesis. Influence of method of synthesis on morphology of nanoparticles, their absorption, diffuse reflectance and emission spectra was observed and studied. Luminescence properties and diffuse reflectance spectra of the sol-gel nanoparticles are also depend on Sm concentrations. At least two types of $Sm^{3+}$ centers were found by emission spectra. These centers have different excitation efficiency by light from the 350 – 450 nm spectral range. Structures of the centers are discussed taking into account crystal structure, possible defects, morphology of the synthesized nanoparticles and their phase compositions.

## 1. Introduction

Vanadate materials are well known and widely used matrices for luminescent rare earth (RE) ions because they satisfy high efficiency of transfer of excitation energy absorbed by matrix to the RE activators [1-5]. The wide range of their applications have attracted significant research efforts to development of new vanadate compositions with improved characteristics depending on requirements of various practical tasks [6-9]. In particular, increase of excitation efficiency under light from near UV and violet spectral ranges is needed for solar energy conversion applications, whereas biological applications require light transformer materials safe to human body with high efficiency of energy transformation in order to decrease x-rays irradiation of living tissues [10-13].

The most attention in many-years investigation of the RE-doped vanadates was paid to study of Eu-activated vanadate compositions [1-4, 12-15]. Indeed, the $Eu^{3+}$ luminescent activators have demonstrated the best efficiency-to-cost rates. At the same time, some task could require improvement of spectroscopy characteristic in despite of cost increase. Besides, investigation of vanadate compositions with various luminescent activators should help to better understand processes of light absorption, energy transfer and radiation emission in vanadate hosts in general. For such purpose, $Sm^{3+}$ ions can be considered as the most promising luminescent activator after the $Eu^{3+}$ ions. Emission spectra of the $Sm^{3+}$ ions are usually observed in the 550 – 700 nm range and can be excited from the 370 – 500 nm range with inner f-f transitions [16-20].

Recently, studying the $Eu^{3+}$-activated $LaVO_4$ vanadate nanoparticles, we have achieved increase of emission intensity in 10 times with various synthesis conditions [21-23]. Question about origin of two types of emission centers in sol-gel $La_{1-x}Eu_xVO_4$ nanoparticles is still open for us. In order to continue our study of formation of additional luminescent centers, in the present work we investigate properties of Sm-activated $LaVO_4$ vanadate nanoparticles. The main attention will be paid to detailed spectroscopy investigation of luminescent properties of the $Sm^{3+}$ ions as possible instrument for clear understanding of structure and origins of their emission centers in the studied matrix.

Synthesis and investigation of the $Eu^{3+}$-doped $LaVO_4$ nanoparticles have been extensively reported, especially at the last decade [21-28]. In despite of this, synthesis and investigation of the $Sm^{3+}$-doped $LaVO_4$ compounds were reported only a few times [22, 29-33]. All these documents have considered $Sm^{3+}$-doped $LaVO_4$ together with other $Ln^{3+}$ dopants. No concentration dependencies have been studied. The reported samples were obtained by organic solvent-solution precipitation reaction [29-31] and by hydrothermal synthesis [32, 33]. Thus, in the present paper we report the first investigation of concentration dependencies of the $Sm^{3+}$-doped $LaVO_4$ nanoparticles synthesized by co-precipitation and sol-gel methods.

## 2. Synthesis of the samples and phase purity
### 2.1. Co-precipitation synthesis

The $La_2O_3$ and $Sm_2O_3$ starting oxides were dissolved into concentrated nitric acids. The stoichiometric ratios of the obtained solutions were mixed homogenously. After 30 min stirring, their pH value was adjusted to about 7.0 - 8.0 with ammonia solution. Then, the $Na_3VO_4$ solution were added into the prepared $La(NO_3)_3$ and $Sm(NO_3)_3$ solutions. The common solutions were stirred to homogeneous states and deposited for few days. After deposition, they were filtered and dried in a drying box. The obtained in such a way precursors were thoroughly homogenized and calcinated in a muffle furnace using step-by-step heating up to temperature 680 °C with 1 hour temperature exposure and intermediate grinding for each 100 °C. Then the samples were heat-treated for 6 hours at 900 °C.

### 2.2. Sol-gel synthesis

Calculated stoichiometric amounts of $La(NO_3)_3$, $Sm(NO_3)_3$ and $NH_4VO_3$ were used as starting reagents. They were taken in necessary quantities and mixed gradually. The solution system was homogenized using a magnetic stirrer. At the next stage, a complexing agent was added to prevent the sedimentation process from the reaction mixture of a similar composition using the deposition method. The citric acid in amount of 5 g was dissolved in a 100 ml of distilled water. Then the solution was heated and added to the reagents solution. This mixture was poured into a graphite cup and placed on a sand bath. The solution gradually evaporated and turned into a gel, and then to a powder. The fine-grained powder was calcined for 5 hours at with a step-by-step heating with 100 °C steps up to 680 °C temperature. More details about methods of synthesis can be found in [23, 34, 35].

### 2.3. Phase compositions

The phase composition and crystal lattice parameters were investigated at Shimadzu LabX XRD-6000 ($Cu_{K\alpha}$-radiation) diffractometer in $10 < 2\Theta < 90°$ angle range with a 2°/min step. The basic samples of the $LaVO_4$ nanoparticles (x = 0) have monoclinic structure for the both applied methods of synthesis [23, 35]. The XRD patterns of the $La_{1-x}Sm_xVO_4$ $0.1 \leq x \leq 0.3$ samples synthesized by co-precipitation and sol-gel methods are shown in Fig. 1. All the investigated $La_{1-x}Sm_xVO_4$ samples synthesized by co-precipitation method are related to tetragonal structure, belonging to the I41/amd space group. The corresponded XRD results are well agreed with the standard card of tetragonal LaVO4 (JCPDS PDF2 32-0504) (see Fig. 1).

It was found that $La_{0.9}Sm_{0.1}VO_4$ sample synthesized by sol-gel method is crystallized in monoclinic structure; a space group is P21/n (see Fig. 1). This result is well matched with a standard card of monoclinic LaVO4 (JCPDS PDF2 50-0367) (see Fig. 1). We see that small quantities of the $Sm^{3+}$ (x ≤ 0.1) ions can enter to the crystal lattice without change of the monoclinic LaVO4 structure. Increasing concentrations of the $Sm^{3+}$ impurities (0.2 ≤ x ≤ 0.3) leads to formation of the monoclinic and tetragonal phases mixture (Fig. 1). The difference in the XRD patterns between monoclinic and tetragonal phases can be clearly distinguished by behavior of the intensive peaks in the regions 23 – 32° and 47-50° of 2θ. When the content of samarium ions increases the relative intensities of the peaks inherent to tetragonal phase (23.79°, 31.95° and 47.32°) increase, while relative intensities of the peaks related to the monoclinic phase (26.20°, 27.81° and 30.14) decrease. These facts point on increase of the tetragonal content with samarium concentration increase.

Thus, proximity of the $La^{3+}$ (1.16 nm) and $Sm^{3+}$ (1.079 nm) ions radii [36] allows common inclusion of the both lanthanum and samarium ions in three plus cations' sites, but doping of $LaVO_4$ basis composition with the $Sm^{3+}$ ions leads to transformation of monoclinic structure on tetragonal with increase of the $Sm^{3+}$ ions. Noted, that concentration range of this transformation strongly depends on method of synthesis: co-precipitation vs sol-gel.

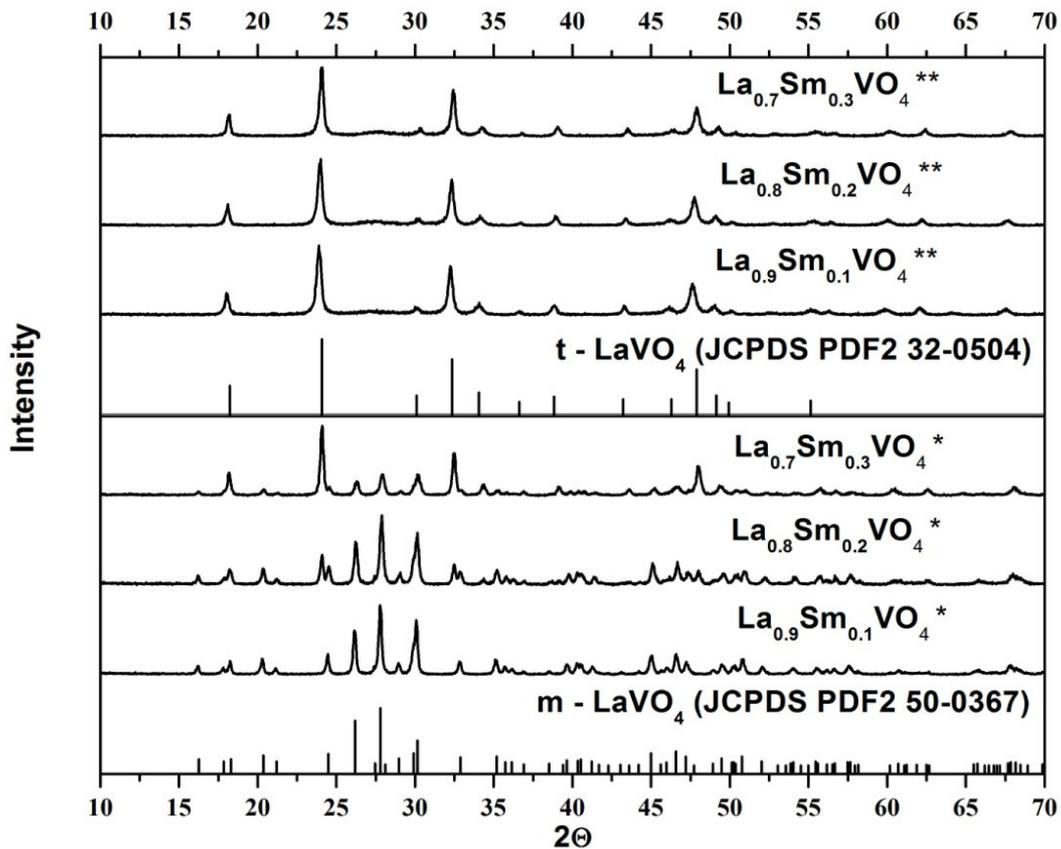

Fig. 1. XRD patterns of $La_{1-x}Sm_xVO_4$ $0.1 \leq x \leq 0.3$ samples synthesized by co-precipitation (**) and sol-gel (*) methods.

## 3. Morphology

Morphology of the synthesized samples was investigated using scanning electron microscope (SEM) Tescan Mira 3 LMU with 20 nm electronic beam diameter. Co-precipitated samples consist of nanoparticles those are slightly agglomerated in grains of different sizes from tens of nanometers to one micron. There are many pores and cavities between nanoparticles in grains. Each of samples consists of nanoparticles of similar shapes and sizes. It is clearly seen from Fig. 2 that the sizes of co-precipitated nanoparticles are increased with increasing of Sm concentration from 20 – 40 nm for $x = 0.1$ to 50 – 70 nm for $x = 0.3$. The shapes of co-precipitated nanoparticles are round, they have forms of distorted cubes without clear corners or distorted spheres. Sol-gel samples consist of nanoparticles those are densely agglomerated in solid grains with sizes about 0.5 – 2.0 microns (Fig. 3). Sizes of nanoparticles are varied in a higher range from 50 to 100 nm for all the samples. It is not observed differences between sizes of nanoparticles with various Sm concentrations for the sol-gel samples. The shapes of sol-gel nanoparticles have polyhedral forms with various proportions: some of particles are near to cubic forms, whereas some of the particles are stretched along one of the axes. Comparing morphology of the nanoparticles synthesized by co-precipitation and sol-gel methods, we should note here that co-precipitation allows to obtain better quality of the nanoparticles, those are characterized by better homogeneity and lower rate of agglomeration.

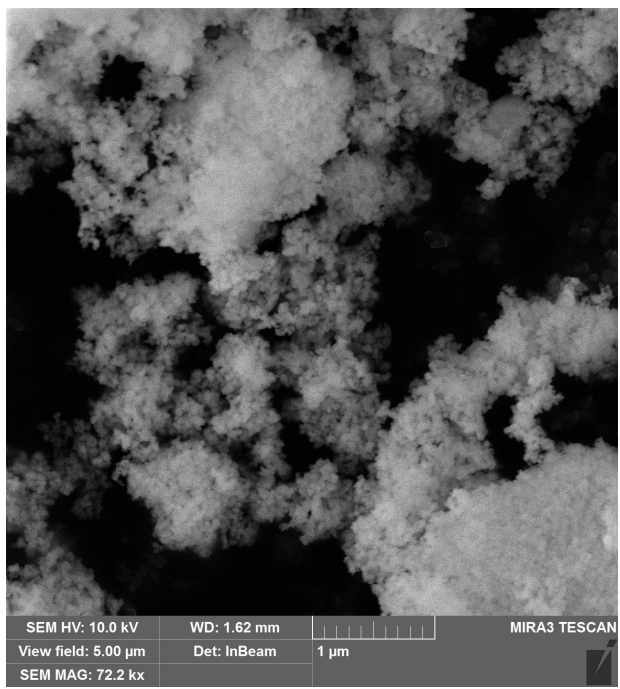 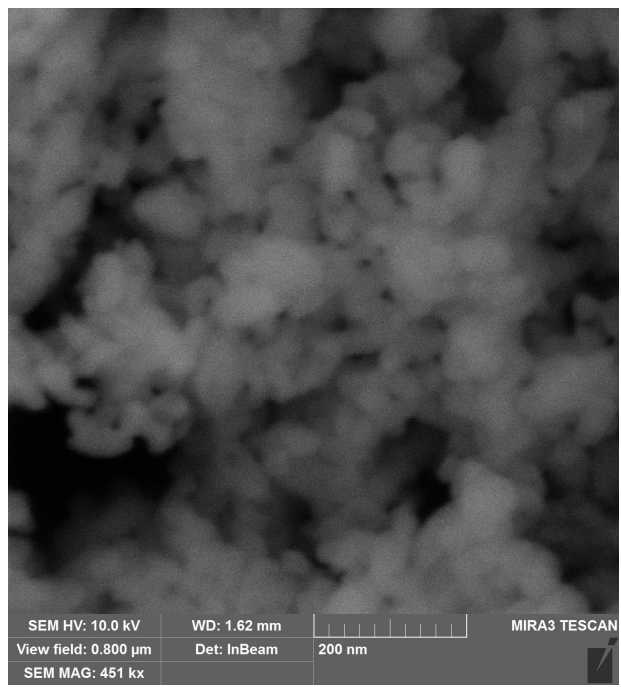

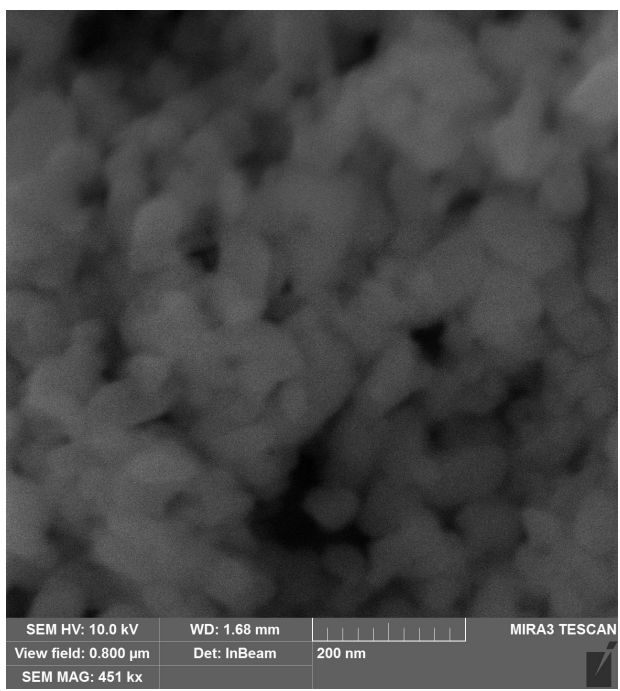 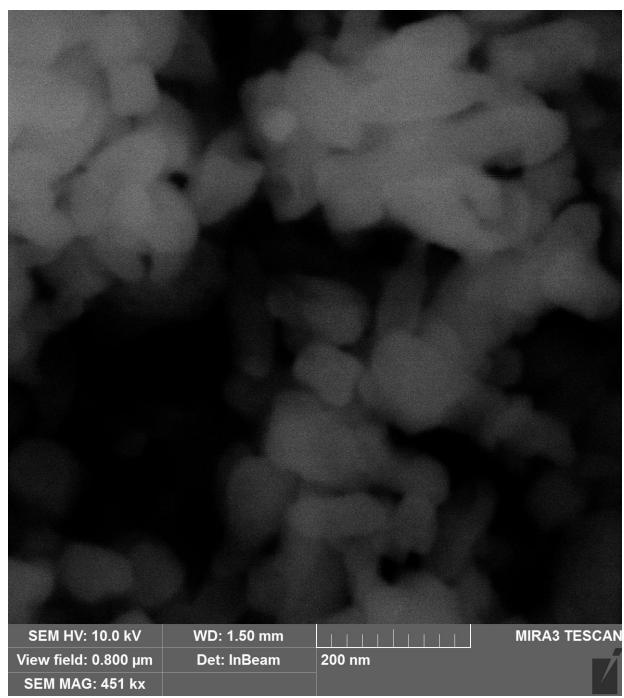

Fig. 2. The SEM images of the La$_{1-x}$Eu$_x$VO$_4$ nanoparticles synthesized by co-precipitation method  x = 0.1 (top), 0.2 (bottom, left), 0.3 (bottom, right).

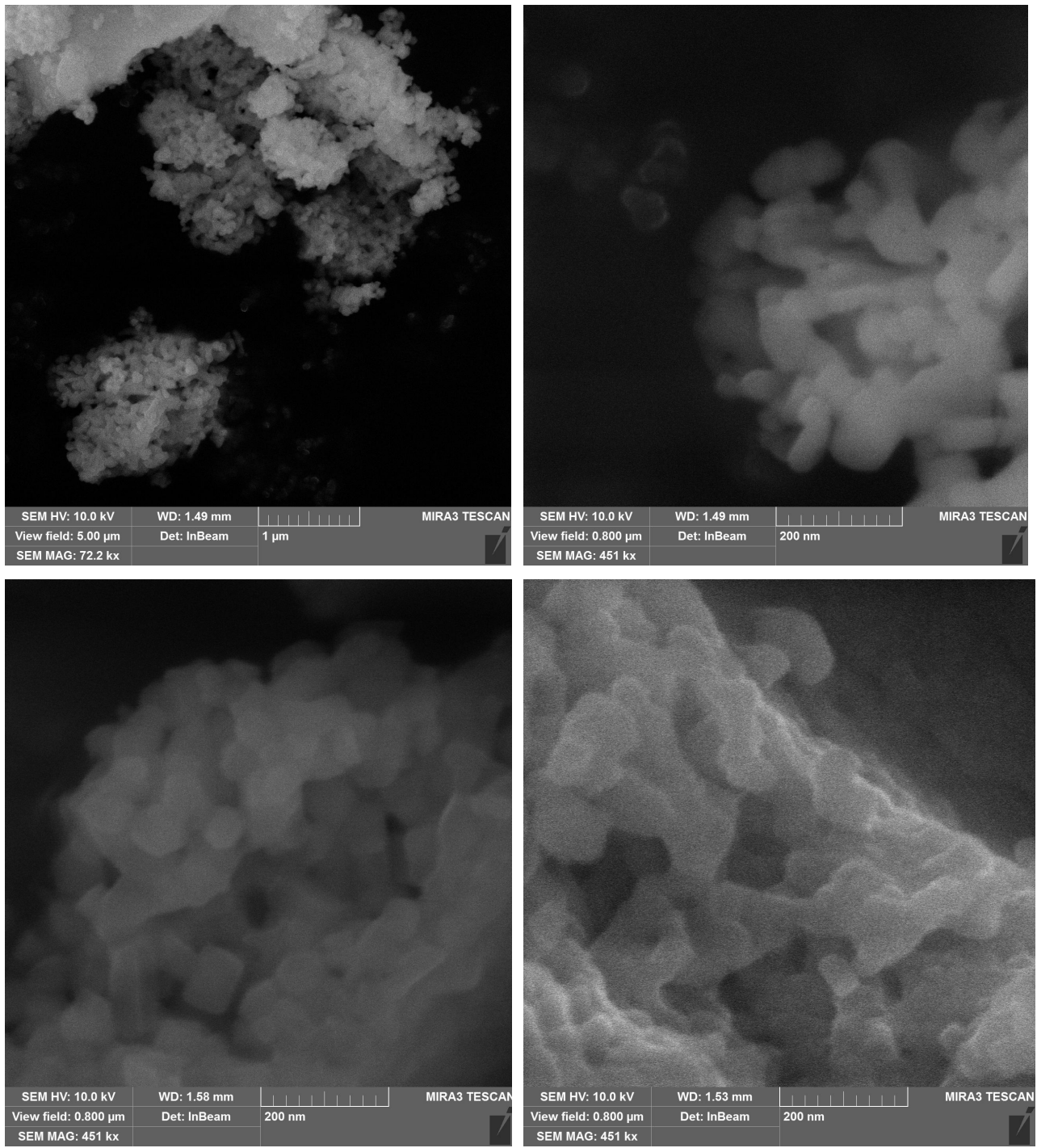

Fig. 3. The SEM images of the $La_{1-x}Eu_xVO_4$ nanoparticles synthesized by sol-gel method x = 0.1 (top), 0.2 (bottom, left), 0.3 (bottom, right).

### 4. IR spectroscopy

IR absorption spectra were measured in the range 400 - 4000 $cm^{-1}$ on PerkinElmer Spectrum BX FTIR spectrometer using the KBr pellet method. The IR spectroscopy investigation is carried out to find differences in the IR spectra of nanoparticles and compare them with differences in crystal structure and luminescence properties.

The measured IR spectra of co-precipitated nanoparticles contain two main peaks near 440 and 800 $cm^{-1}$ (Fig. 4). These peaks are corresponded to bending and stretching vibrations of O–V–O bonds of the

$VO_4^{3-}$ vanadate anion. The peaks at 440 cm$^{-1}$ and stronger wide band in the range 700–1000 cm$^{-1}$ with peaks around 800 cm$^{-1}$ should be assigned, respectively, to $\nu_4$ and $\nu_3$ vibration modes of the $VO_4^{3-}$ anion [37-39]. The IR absorption of nanoparticles synthesized by different methods is characterized by some spectral differences. The spectra of the co-precipitated nanoparticles don't depend on the $Sm^{3+}$ concentration, whereas spectra of the sol-gel nanoparticles depend on concentration. The peak at 440 cm$^{-1}$ that is corresponded to the $\nu_4$ deformation modes has strongly increased intensity for the sol-gel nanoparticles (near 5 times at x = 0.1) and its intensity decreases with increasing of Sm concentration in the sol-gel nanoparticles (Fig. 4, right). The main bands in the range 700–1000 cm$^{-1}$ corresponded to the $\nu_3$ mode for the co-precipitated nanoparticles are characterized by one peak at 794 cm$^{-1}$. In the spectra of the sol-gel nanoparticles this band is characterized by four peaks at 775, 794, 825 and 853 cm$^{-1}$.

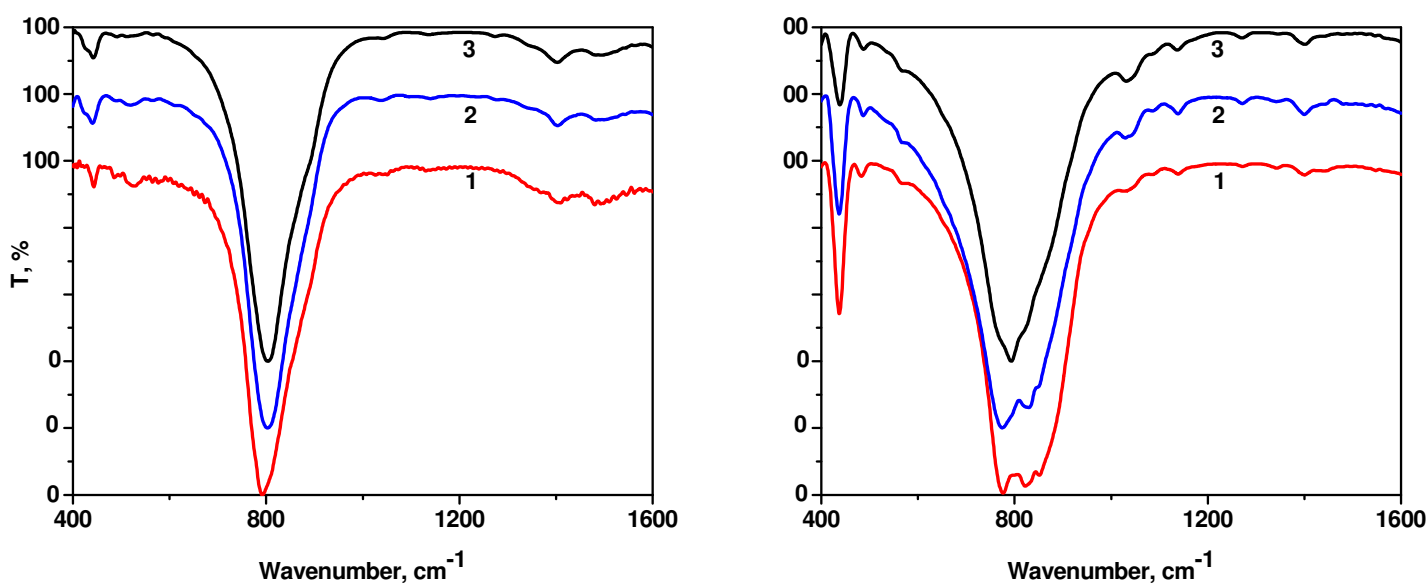

Fig. 4. FTIR absorption spectra of the $La_{1-x}Sm_xVO_4$ samples synthesized by co-precipitation (left) and sol-gel (right) methods, x = 0.1 (1), 0.2 (2) and 0.3 (3).

## 5. Optical spectroscopy

Diffuse reflectance spectra were investigated in 250 – 2500 nm range using Perkin Elmer Lambda 950 spectrometer. Reflectance spectra of all the samples contain sharp edge with 50 % reflection at 340 nm. Visible part of the spectra contains also narrow spectral failures peaking at 365, 380, 407 (the strongest), 421, 442, 458, 467, 479, 490, 501 and 532 nm (Fig. 5). Intensities of these peaks are increased with Sm concentration and strongly depend on method of synthesis: the peaks are considerably stronger for the sol-gel samples. By their positions these peaks can be assigned to f-f transitions in the $Sm^{3+}$ ions from $^6H_{5/2}$ level on various higher electron levels [40-42]. The peak assignments are collected in Table 1. The near IR part of the reflectance spectra contains peaks of the $^6H_{5/2} \rightarrow {}^6F_j$ transitions. Their positions are also collected in Table 1.

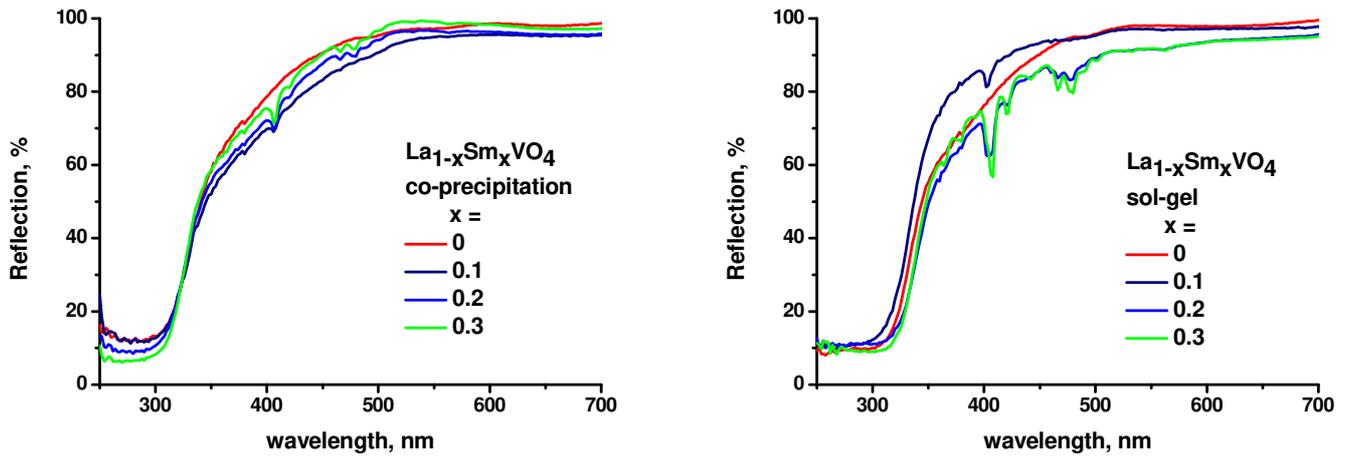

Fig. 5. UV and Visible part of reflectance spectra of the $La_{1-x}Sm_xVO_4$ samples synthesized by co-precipitation (left) and sol-gel (right) methods, x = 0, 0.1, 0.2 and 0.3.

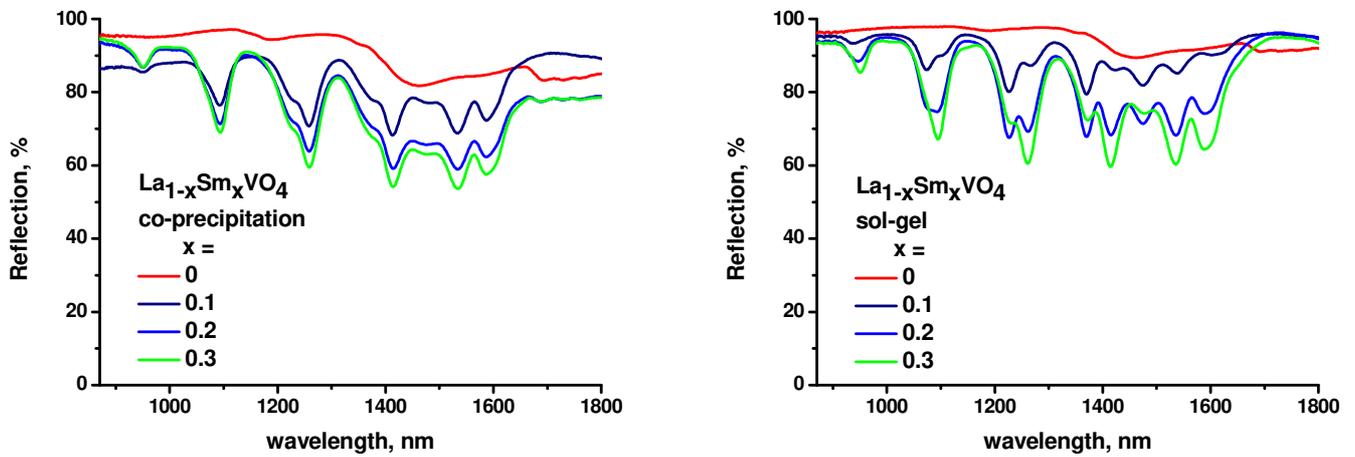

Fig. 6. Near IR part of reflectance spectra of the $La_{1-x}Sm_xVO_4$ samples synthesized by co-precipitation (left) and sol-gel (right) methods, x = 0, 0.1, 0.2 and 0.3.

Table 1. Assignments of peaks observed in the reflectance spectra of the $La_{1-x}Sm_xVO_4$ samples

| Peak positions | Transitions | Comments | Peak positions, nm | Transitions | Comments |
|---|---|---|---|---|---|
| 365 | $^6H_{5/2} \to {}^4D_{3/2}$ | | 934 | $^6H_{5/2} \to {}^6F_{11/2}$ | |
| 380 | $^6H_{5/2} \to {}^6P_{7/2}$ | | 950 | | |
| 407 | $^6H_{5/2} \to {}^4K_{11/2}$ | the strongest | 1075 | $^6H_{5/2} \to {}^6F_{9/2}$ | |
| 421 | $^6H_{5/2} \to {}^6G_{9/2}$ | strong | 1095 | | |
| 442 | $^6H_{5/2} \to {}^4L_{13/2}$ | | 1104 | | strong |
| 458 | $^6H_{5/2} \to {}^4M_{17/2}$ | | 1225 | $^6H_{5/2} \to {}^6F_{7/2}$ | |
| 467 | $^6H_{5/2} \to {}^4F_{5/2}$ | strong | 1260 | | |
| 479 | $^6H_{5/2} \to {}^4I_{13/2}$ | strong | 1268 | | strong |
| 490 | $^6H_{5/2} \to {}^4I_{11/2}$ | | 1368 | $^6H_{5/2} \to {}^6F_{5/2}$ | |

| 501 | $^6H_{5/2} \to {}^4M_{15/2}$ | | 1414 | | strong |
| 532 | $^6H_{5/2} \to {}^4G_{7/2}$ | | 1475 | $^6H_{5/2} \to {}^6F_{3/2}$ | |
| | | | 1538 | | strong |
| | | | 1590 | $^6H_{5/2} \to {}^6F_{1/2}$ | strong |

## 6. Luminescence spectroscopy

Luminescent spectra were measured using high resolution DFS-12 (LOMO) diffraction spectrometer equipped with FEU-79 photomultiplier. Measured emission spectra consist of four groups of lines in 550 – 725 nm spectral range. There are groups located near 550 – 580, 580 – 620, 625 – 670 and 680 – 725 nm. The spectra of our samples synthesized by co-precipitation and sol-gel methods have distinctive differences (Fig. 7). Spectra of the samples synthesized by sol-gel method have linear structure much more reach for details and depend on concentration (Fig. 7, right). Spectra of the samples synthesized by co-precipitation method have less of details and don't depend on concentration (Fig. 7, left). Differences between the spectra are the best seen for the $La_{0.9}Sm_{0.1}VO_4$ samples (Fig. 7, curves 1). The most intensive emission peak for the co-precipitated samples is located in the 625 – 670 nm spectral range at 646 nm. The most intensive peak for the sol-gel samples is located in the 580 – 620 nm range at 597 nm. This difference is observed for all applied excitations (Fig. 8). The lines observed in the emission spectra of the investigated $La_{1-x}Sm_xVO_4$ nanoparticles are corresponded to well-known inner f-f transitions in the $Sm^{3+}$ ions. Groups of lines located in the 550 – 580, 580 – 620, 625 – 670 and 680 – 725 nm spectral ranges are caused by the $^4G_{5/2} \to {}^6H_{5/2}$, $^6H_{7/2}$, $^6H_{9/2}$ and $^6H_{11/2}$ transitions, respectively [17-20]. Positions of the observed lines and their assignments are collected in Table 2.

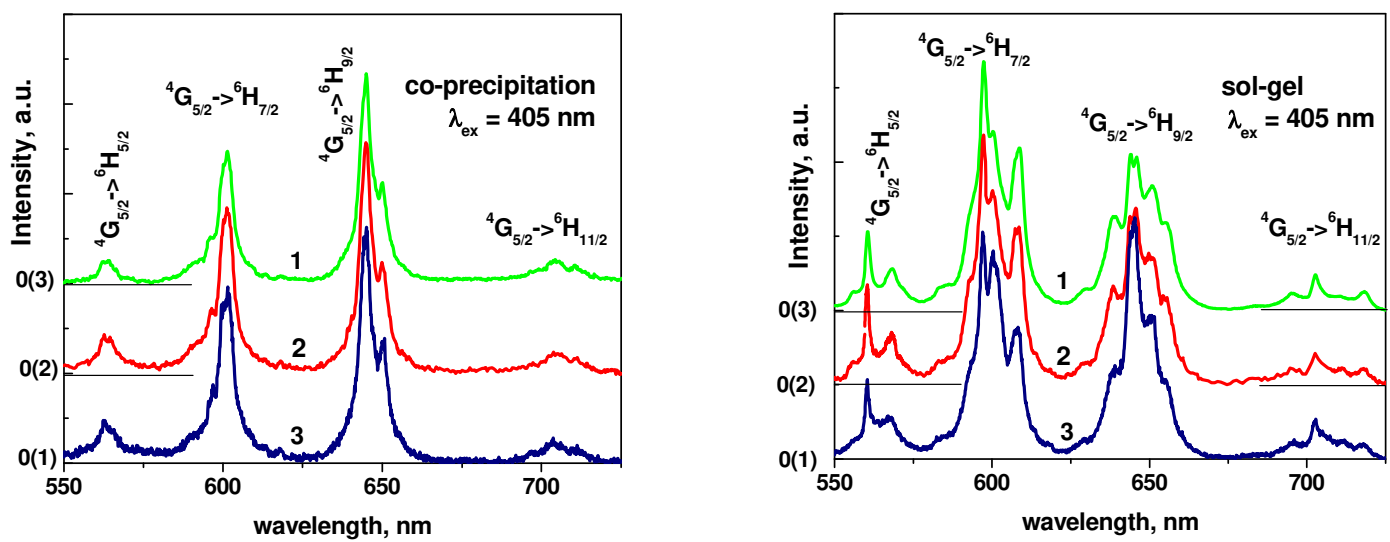

Fig. 7. Emission spectra of the $La_{1-x}Sm_xVO_4$ samples synthesized by co-precipitation (left) and sol-gel (right) methods, x = 0.1 (1), 0.2 (2), 0.3 (3), $\lambda_{ex}$ = 405 nm.

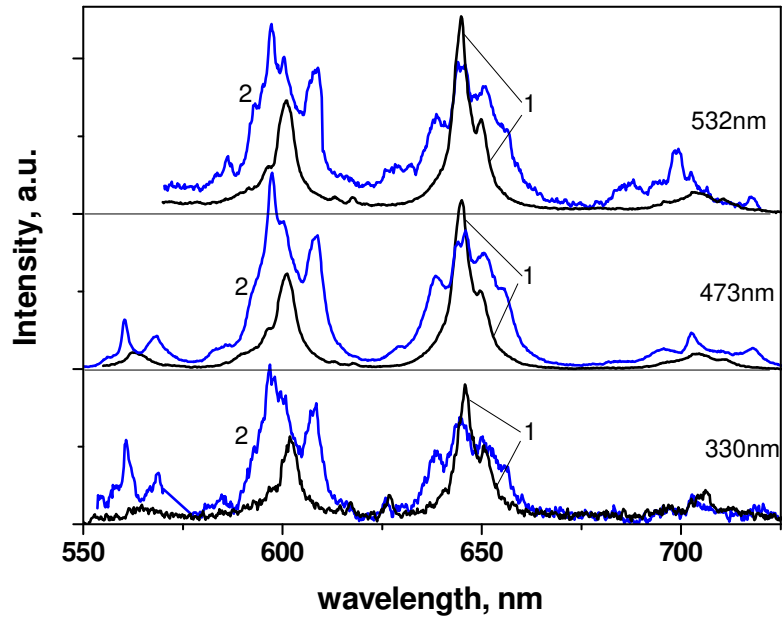

Fig. 8. Emission spectra of the $La_{0.9}Sm_{0.1}VO_4$ samples synthesized by co-precipitation (1) and sol-gel (2) methods; $\lambda_{ex}$ = 532, 473, 330 nm.

Table 2. Peak positions and their assignments of all the emission peaks observed in the spectra of the $La_{1-x}Sm_xVO_4$ samples synthesized by co-precipitation and sol-gel methods measured at various excitations

| Transition | Type | Range, nm | Peak positions for co-precipitated samples, nm | Peak positions for sol-gel samples, nm |
|---|---|---|---|---|
| $^4G_{5/2} \rightarrow {}^6H_{5/2}$ | magnetic dipole | 550 – 580 | 562, 566, 572 | 556, 560*, 568, 573 |
| $^4G_{5/2} \rightarrow {}^6H_{7/2}$ | electric and magnetic dipole | 580 – 620 | 590, 596, 599, 601*, 608, 613, 618 | 583, 593, 597**, 601, 604, 608, 609, 617 |
| $^4G_{5/2} \rightarrow {}^6H_{9/2}$ | electric dipole | 625 – 670 | 629, 636, 639, 646**, 650, 655 | 629, 635, 638, 644*, 646*, 649, 651, 655, 663, 667 |
| $^4G_{5/2} \rightarrow {}^6H_{11/2}$ | electric dipole | 680 – 725 | 682, 696, 704, 711, 716 | 696, 702, 712, 719 |

\* - strong

\*\* - the strongest

The most intensive $^4G_{5/2} \rightarrow {}^6H_{7/2}$ and $^4G_{5/2} \rightarrow {}^6H_{9/2}$ transitions correspond to the orange and red luminescence, respectively. These transitions have different nature that is noted in Table 2. The ratio of the transitions intensity (O/R ratio) gives more information about covalence and symmetry of the nearest surrounding of the $Sm^{3+}$ ions (Table 3). The $^4G_{5/2} \rightarrow {}^6H_{7/2}$ is mixed magnetic dipole and electric dipole transition. The $^4G_{5/2} \rightarrow {}^6H_{9/2}$ is purely electric dipole transition. It is known if intensity of magnetic dipole transition is higher, then the lower symmetry of crystal surrounding of the $Sm^{3+}$ ions takes place [17, 42-44]. Fig. 7 shows that emission spectra of the samples synthesized by co-precipitation don't depend on Sm concentration whereas intensity distributions of lines in the emission spectra of sol-gel nanoparticles depend on concentration. Fig. 9 shows detailed views of emission of the sol-gel samples in the spectral ranges corresponded to the most intensive $^4G_{5/2} \rightarrow {}^6H_{7/2}$ and $^4G_{5/2} \rightarrow {}^6H_{9/2}$ transitions. It is possible to clearly see decrease of contributions of emission peaks at 597, 609 and 644 nm with increasing of Sm concentration if compare intensities of these lines with neighboring peaks at 601, 607 and 646 nm, respectively (Fig. 9).

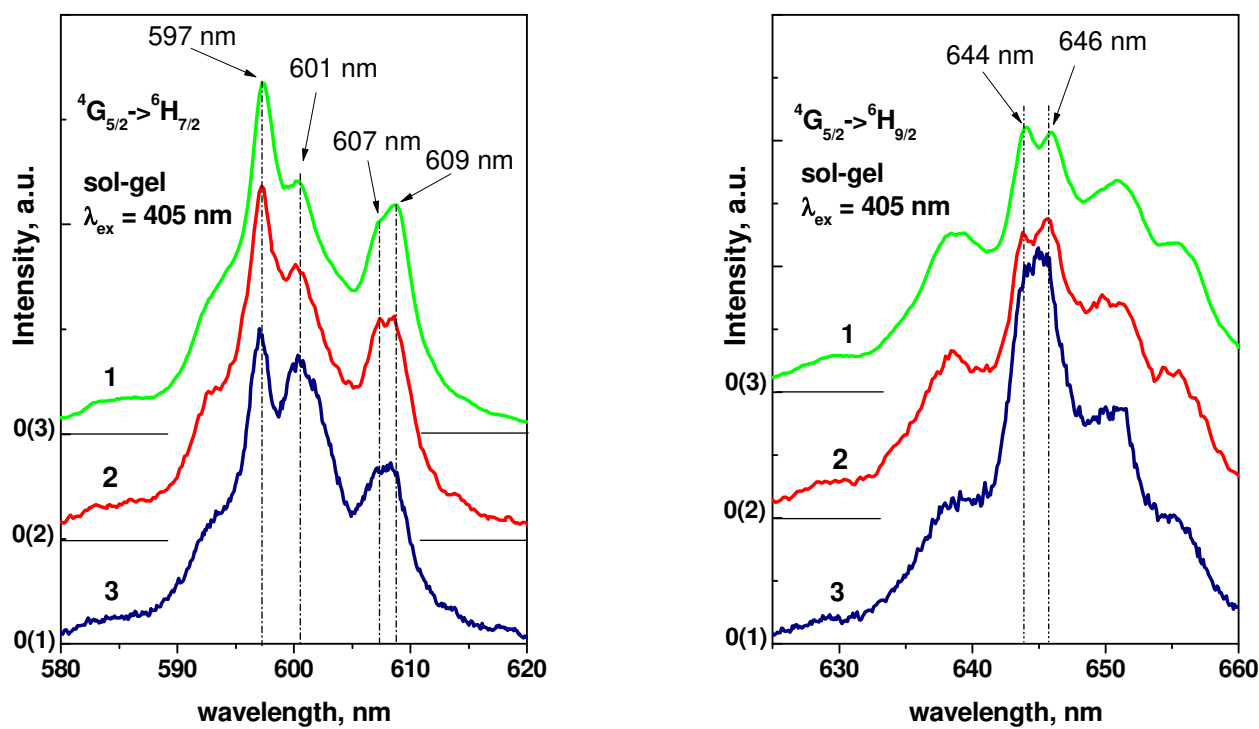

Fig. 9. Detailed ranges of the $^4G_{5/2} \rightarrow {}^6H_{7/2}$ (left) and $^4G_{5/2} \rightarrow {}^6H_{9/2}$ (right) transitions in the emission spectra of the $La_{1-x}Sm_xVO_4$ samples synthesized by sol-gel method, x = 0.1 (1), 0.2 (2), 0.3 (3), $\lambda_{ex}$ = 405 nm.

Table 3. The O/R ratios in the emission spectra of the $La_{1-x}Sm_xVO_4$ samples synthesized by co-precipitation and sol-gel methods for various concentrations, x, excitation 405 nm

| Method of synthesis | Concentration, x | O/R ratio |
|---|---|---|
| co-precipitation | 0.1 | 0.6 |
|  | 0.2 | 0.6 |
|  | 0.3 | 0.75 |
| sol-gel | 0.1 | 1.6 |
|  | 0.2 | 1.4 |
|  | 0.3 | 0.95 |

## 7. Excitation

Excitation spectra were investigated using Xenon lamp source DKsL-1000 Wt and DMR-4 double prism monochromator. To take into account nonlinear dispersion of the excitation monochromator, the measured spectra were converted into energy scale using $\lambda^2$ correction factor. The obtained spectra are given below in the energy scale, but the wavelength scales are also marked at the top of Figures in order to continue discussion of the luminescence properties using wavelength scale as above.

Excitation spectra consist of broad band in 250 – 350 nm spectral range with maxima near 320 nm and three groups of more narrow and structured spectral components. There are groups in 350 – 380 (weak), 385 – 440 and 450 – 520 nm spectral ranges (Fig. 10). Each of these groups contains several spectral peaks. The spectra of the samples synthesized by different methods have spectral differences those are the same for various Sm concentrations. Maxima of the broad band is shifted in a long wave length range from 320 nm for the co-precipitated samples to 330 nm for the sol-gel samples, whereas maxima of the narrow structures bands are shifted in a short wavelength range for the sol-gel samples (Figs. 10, 11). At the same time, spectra registered for the sol-gel samples in the lines with different behavior (see Fig. 9) are characterized by the same spectral distribution (Fig. 11). Relative contribution of the 320 nm band also depends on method of synthesis. It is considerably higher for the samples synthesized by co-precipitation method.

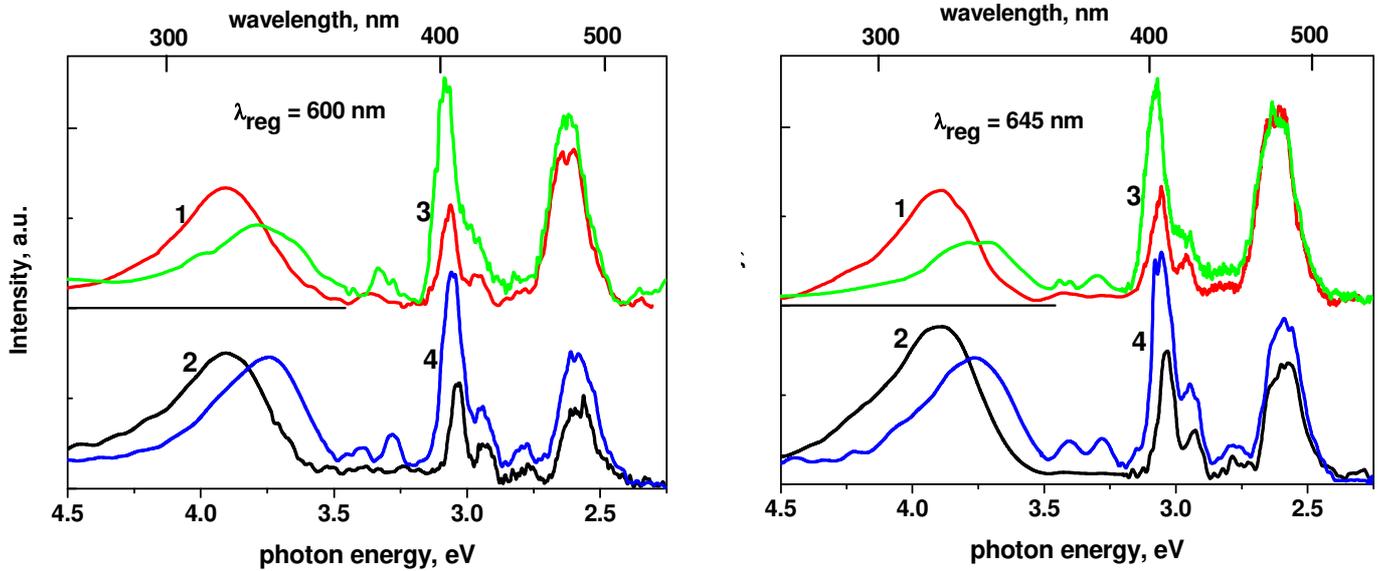

Fig. 10. Excitation spectra of the $La_{1-x}Sm_xVO_4$ samples synthesized by co-precipitation (1, 2) and sol-gel (3, 4) methods, x = 0.2 (1, 3), 0.3 (2, 4), $\lambda_{reg}$ = 600 (left) and 645 nm (right).

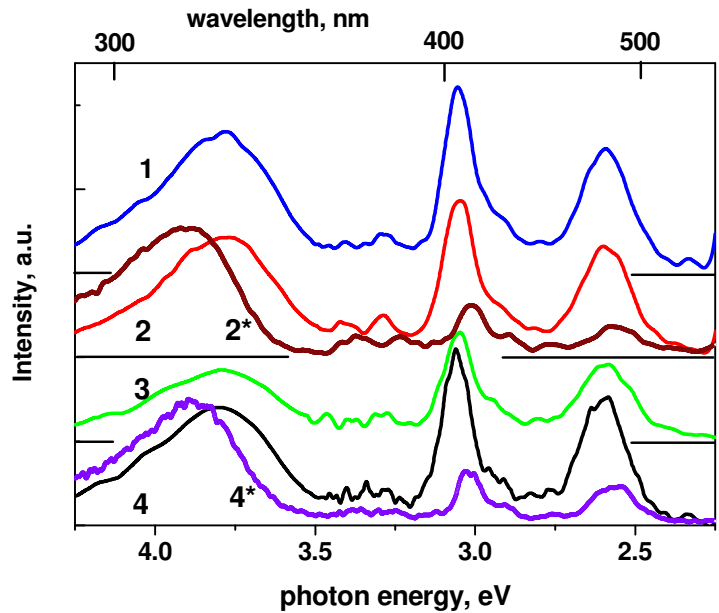

Fig. 11. Excitation spectra of the $La_{0.9}Sm_{0.1}VO_4$ samples synthesized by co-precipitation (marked by *) and sol-gel methods, $\lambda_{reg}$ = 597 (1), 600 (2), 639 (3) and 645 nm (4).

The observed excitation spectra components have various origins. The broad band in 250 – 350 nm spectral range should be assigned to transitions in the $VO_4^{3-}$ groups [45-47], the maxima observed in the 350 – 380, 385 – 440 and 450 – 520 nm spectral ranges should be assigned to inner f-f transitions in the $Sm^{3+}$ ions. There are $^6H_{5/2} \to {}^4D_{3/2}$ and $^6H_{5/2} \to {}^6P_{7/2}$ transitions in the 350 – 380 nm spectral range, $^6H_{5/2} \to {}^4K_{11/2}$, $^6H_{5/2} \to {}^6G_{9/2}$ and $^6H_{5/2} \to {}^4L_{13/2}$ transitions in the 385 – 440 nm spectral range and $^6H_{5/2} \to {}^4M_{17/2}$, $^6H_{5/2} \to {}^4F_{5/2}$, $^6H_{5/2} \to {}^4I_{13/2}$, $^6H_{5/2} \to {}^4I_{11/2}$ transitions in the 450 – 520 nm spectral range [41-44].

## 8. Discussion

In a general case, the observed differences between spectral properties of the samples obtained by different methods and concentration dependencies of the spectral properties should be discussed in connection with phase compositions and morphology of the samples. But, in this study we have not found essential differences in morphology of the samples with various Sm concentrations, especially it concerns samples obtained by sol-gel synthesis. At the same time, we have observed distinctive changes with concentration of Sm ions in the sol-gel samples at all types of the measured spectra. Thus, we should pay the main attention to study correlations of changes of spectral properties with the observed changes of crystal phase structures of the synthesized samples (Fig. 1).

As it was shown above in the section 2.3, all the co-precipitated samples have single phase structure with tetragonal crystal lattice, whereas sol-gel samples have two phase structure. Moreover, contributions of monoclinic and tetragonal phases in the synthesized sol-gel samples strongly depend on Sm concentrations. The observed phase compositions could effects on spectral properties of the synthesized vanadate nanoparticles [25, 48, 49]. At tetragonal $LaVO_4$ crystal lattice the La and V cations are located at high-symmetry positions. The V atoms are coordinated by four equivalent oxygen atoms forming the perfect $VO_4$ tetrahedrons with $T_d$ local symmetry of oxygen surrounding. The La atoms are coordinated by eight oxygen atoms with identical four short bond lengths and four long bond lengths with $D_{2d}$ local symmetry of oxygen surrounding [37, 48]. At monoclinic crystal structure, the V and La atoms are four- and nine-fold coordinated, respectively, with $C_s$ local symmetry of oxygen surrounding for the both cations. Thus, different symmetries of local surrounding are expected to have distinctive effects on spectral properties of both $Sm^{3+}$ ions and $VO_4^{3-}$ groups in the monoclinic and tetragonal crystal lattices.

Indeed, the observed differences in the IR spectra of nanoparticles synthesized by co-precipitation and sol- gel methods can be explained taking into account the above described results of XRD analysis. The $v_3$ stretching vibration mode of the $VO_4^{3-}$ group is triple degenerated. It is not split in the spectra of the samples containing $VO_4^{3-}$ groups of tetrahedral symmetry (tetragonal crystal lattice, nanoparticles synthesized by co-precipitation) and it is split on three components in the spectra of the samples containing $VO_4^{3-}$ groups of lowered symmetry ($C_s$ symmetry, monoclinic crystal lattice) [37, 48]. Observation of four peaks in the range of $v_3$ vibration mode for the sol-gel nanoparticles can be caused by presence of two phases in their compositions – tetragonal (one component at 794 $cm^{-1}$), and monoclinic – three components at 775, 825 and 853 $cm^{-1}$. Increased intensity of the $v_4$ deformation mode in the spectra of sol-gel samples is also can be ascribed to features of the monoclinic crystal lattice. Such assumption is supported by the observed decrease of relative intensity of the corresponded peak with increasing of Sm concentration in the sol-gel samples: according to XRD analysis, relative content of tetragonal phase increases in the sol-gel samples with increasing of the Sm concentration (Fig. 1).

Some observed features of fine structure of the reflection spectra are also explained by different crystal phase compositions of the co-precipitated and sol-gel nanoparticles. It is clearly seen from Fig. 6, that intensity distributions of the observed peaks don't depend on Sm concentration for the samples synthesized by co-precipitation and strongly depend on concentration for the sol-gel samples. The IR reflectance spectrum of the sol-gel samples with x = 0.3 is different from the spectra of the samples with x = 0.2 and 0.1. At the same time, this spectrum is similar to the corresponded spectra measured for the samples synthesized by co-precipitation. Taking into account results of the XRD study we should make an assumption that changes of intensity distributions between peaks in the near IR part of reflectance spectra of the $La_{1-x}Sm_xVO_4$ nanoparticles are correlated with their crystal phase compositions: tetragonal or monoclinic.

The observed in the emission spectra two types of lines those are characterized by different behavior with Sm concentration in the sol-gel samples and total differences between structures of emission spectra of the samples obtained with different methods can be connected with tetragonal and monoclinic crystal phases in these samples. In such a case two sets of emission lines with different behavior are connected with $Sm^{3+}$ ions arranged in tetragonal crystal lattice (peaks at 601, 607, 646 nm and all peaks herein listed in Table 2 for the co-precipitated samples) and monoclinic crystal lattice (peaks at 597, 609, 644 nm and other peaks observed for the sol-gel samples only). This assumption agrees with estimations of the O/R rates and covalency of the nearest surrounding of the $Sm^{3+}$ ions. According to Table 3, the $Sm^{3+}$ ions in sol-gel nanoparticles are incorporated in sites with a stronger covalent bonding with the nearest oxygen surrounding that is characteristic for more asymmetric arrangement. The observed decrease of the asymmetric rate with concentration is connected with contribution of the tetragonal phase in the total emission spectra.

At the same time it should be noted here that contribution of transitions occurred in the $Sm^{3+}$ ions arranged monoclinic phase is still dominated even for emission spectra of the $La_{1-x}Sm_xVO_4$ sol-gel nanoparticles with x = 0.3 where according to XRD analysis content of tetragonal phase significantly prevail content of the monoclinic phase (Fig. 1). It was shown above that results of XRD analysis are in good agreement with the results of diffuse reflectance measurements where spectra of the sol-gel nanoparticles with x = 0.3 are the same with spectra of the tetragonal co-precipitated nanoparticles. The observed dissimilarity between behavior of characteristic of the sol-gel nanoparticles with Sm concentration means that probabilities of the luminescent transitions in the $Sm^{3+}$ ions arranged monoclinic phase are much higher than ones for the tetragonal phase. Therefore, in order to achieve high intensity of luminescence of the $Sm^{3+}$ ions on the orthovanadate host one should satisfy monoclinic structure of its lattice.

As for fine structure of f-f transitions in the excitation spectra, it is not possible to clearly separate peaks of all the contributing transitions. We can only suppose that spectral differences in the intensities of f-f transitions observed for the samples synthesized by different methods are connected with above described high and low symmetries of the nearest surrounding of the $Sm^{3+}$ ions in the tetragonal and monoclinic crystal lattices, respectively. It is obviously, that samples with low symmetry monoclinic structure are characterized with higher contributions of the f-f transitions in the excitation spectra.

The broad excitation band in 250 – 350 nm spectral range that is caused by transitions in the $VO_4^{3-}$ groups. Its contribution is higher for the co-precipitated samples with tetragonal structure. This can be evidence of a higher rate of transfer of excitation energy from the $VO_4^{3-}$ groups to the $Sm^{3+}$ ions in tetragonal crystal lattice. Shift of maxima position of this band dependently on method of synthesis can be caused by its complex character. It is well known that excitation bands of vanadate compounds in the 270 – 350 nm spectral range are caused by $^1A_1 \rightarrow {}^1T_2$ and $^1A_1 \rightarrow {}^1T_1$ transitions in the $VO_4^{3-}$ anions [45-47]. The latter transition is forbidden by symmetry rules and it cannot be observed in the spectra of undistorted $VO_4^{3-}$ tetrahedrons those are part of the tetragonal crystal lattice of vanadates (the co-precipitated samples). In the monoclinic crystal lattice (sol-gel samples) the distorted $VO_4^{3-}$ groups have $C_s$ symmetry that cancels symmetry forbiddance. Thus, shift of maxima positions of the broad excitation band from 320 nm to 330 nm should be assigned to contributions of additional spectral components in the total spectra of the sol-gel samples. But, arising of additional spectral components in excitation of the samples with monoclinic crystal structure don't satisfy better transfer of absorbed excitation energy from the $VO_4^{3-}$ groups to the $Sm^{3+}$ luminescent centers as the 270 – 350 nm excitation band is characterized by essentially higher intensity for the samples with tetragonal crystal structure.

Therefore, crystal structure of the $La_{1-x}Sm_xVO_4$ nanoparticles defines what types of electron transitions will give a higher contribution in excitation of the $Sm^{3+}$ luminescence. Transitions in the $VO_4^{3-}$ groups give higher contribution for the samples with tetragonal structure whereas f-f transitions in $Sm^{3+}$ ions give higher contribution for the samples with monoclinic structure. Taking into account that peaks of these two types of transitions are located in the UV and visible spectral ranges, respectively, we should make a conclusion that crystal structure of the vanadate nanoparticles effects on efficiency of luminescent transformation from the UV or visible spectral ranges.

## 9. Conclusions

The $La_{1-x}Sm_xVO_4$ (0.1 ≤ x ≤ 0.3) nanoparticles were synthesized by co-precipitation and sol-gel methods. Phase compositions of the samples depend on method of synthesis and Sm concentration. The samples synthesized by co-precipitation are characterized by tetragonal structure for all the investigated samples. The sol-gel samples are characterized by monoclinic structure for 0.1 ≤ x ≤ 0.2 and two phases (tetragonal and monoclinic) for 0.2 ≤ x ≤ 0.3.

SEM images, IR absorption spectra, reflectance spectra and luminescence properties of the synthesized nanoparticles were investigated.

Method of synthesis influences on the morphology of the investigated nanoparticles and their spectral properties.

Emission of the investigated samples is observed in the 500 – 725 nm spectral range and it consists of narrow spectral lines. These lines are caused by the $^4G_{5/2} \to {}^6H_J$ electron transitions in the $Sm^{3+}$ ions. Groups of lines located in the 550 – 580, 580 – 620, 625 – 670 and 680 – 725 nm spectral ranges are assigned to the $^4G_{5/2} \to {}^6H_{5/2}, {}^6H_{7/2}, {}^6H_{9/2}$ and ${}^6H_{11/2}$ transitions, respectively.

Excitation spectra consist of the broad band in 250 – 350 nm spectral range that is caused by transitions in the $VO_4^{3-}$ groups and narrow bands in the 350 – 380, 385 – 440 and 450 – 520 nm spectral ranges caused by the f-f transitions in the $Sm^{3+}$ ions. Crystal structure of the $La_{1-x}Sm_xVO_4$ nanoparticles defines what types of electron transitions give a higher contribution in excitation of the $Sm^{3+}$ luminescence.

The assumption is made that all the observed differences in spectral properties of the nanoparticles synthesized by different methods can be explained by their phase composition. Crystal structure of the vanadate nanoparticles influences significantly on efficiency of luminescent transformation of light from the UV and violet spectral ranges.

**Acknowledgments**

This project has received funding from Ministry of Education and Science of Ukraine and from the EU-H2020 research and innovation program under grant agreement No 654360 having benefited from the access provided by Institute of Electronic Structure & Laser (IESL) of Foundation for Research & Technology Hellas (FORTH) in Heraklion, Crete, Greece within the framework of the NFFA-Europe Transnational Access Activity.

**References**

1. F.C. Palilla, A.K. Levine, M. Rinkevics, Rare Earth Activated Phosphors Based on Yttrium Orthovanadate and Related Compounds, J. Electrochem. Soc., 1965, v. 112(8), p. 776-779.
2. G. Panayiotakis, D. Cavouras, I. Kandarakis, C. Nomicos, A study of X-ray luminescence and spectral compatibility of europium-activated yttrium-vanadate ($YVO_4$:Eu) screens for medical imaging applications, Appl. Phys. A., 1996, v. 62(5), p. 483-486.
3. Jong Hyuk Kang, Won Bin Im, Dong Chin Lee, Jin Young Kim, Duk Young Jeon, Yun Chan Kang, Kyeong Youl Jung, Correlation of photoluminescence of $(Y,Ln)VO_4:Eu^{3+}$ (Ln = Gd and La) phosphors with their crystal structures, Solid State Communications, 2005, v. 133, p. 651-656.
4. V.B. Bhatkar, Synthesis and Luminescence Properties of Yttrium Vanadate based Phosphors, International Journal of Engineering Science and Innovative Technology, 2013, v. 2, p. 426-432.
5. A.H. Krumpel, E. van der Kolk, E. Cavalli, P. Boutinaud, M. Bettinelli, P. Dorenbos, Lanthanide 4f-level location in $AVO_4:Ln^{3+}$ (A = La, Gd, Lu) crystals, J. Phys.: Condens. Matter., 2009, v. 21, p. 115503-8.
6. Qianming Wang, Zhengyang Zhang, Yuhui Zheng, Weisheng Cai and Yifei Yu, Multiple irradiation triggered the formation of luminescent $LaVO_4$: $Ln^{3+}$ nanorods and in cellulose gels,

Cryst. Eng. Comm., 2012, v. 14, p. 4786-4793.

7. Zhenhe Xu, Chunxia Li, Zhiyao Hou, Chong Peng and Jun Lin, Morphological control and luminescence properties of lanthanide orthovanadate LnVO$_4$ (Ln = La to Lu) nano-/microcrystals via hydrothermal process, Cryst. Eng. Comm., 2011, v. 13, p. 474–482.

8. Kai Li, Rik Van Deun, Eu$^{3+}$/Sm$^{3+}$-doped Na$_2$BiMg$_2$(VO$_4$)$_3$ from substitution of Ca$^{2+}$ by Na$^+$ and Bi$^{3+}$ in Ca$_2$NaMg$_2$(VO$_4$)$_3$: Color-tunable luminescence via efficient energy transfer from VO$_4^{3-}$ to Eu$^{3+}$/Sm$^{3+}$ ions, Dyes and Pigments, 2018, v. 155, p. 258–264.

9. Y. Zhu, Y. Wang, J. Zhu, D. Zhou, D. Qiu, W. Xu, X. Xu, Z. Lu, Plasmon multiwavelength-sensitized luminescence enhancement of highly transparent Ag/YVO$_4$:Eu3+/PMMA film, J. of Luminescence, 2018, v. 200, p. 158-163.

10. V. Kumar, A.F. Khan, S. Chawla, Intense red-emitting multi-rare-earth doped nano-particles of YVO 4 for spectrum conversion towards improved energy harvesting by solar cells, J. Phys. D: Appl. Phys., 2013, v. 46, p. 365101-9.

11. S.G. Nedilko, V. Chornii, O. Chukova, V. Degoda, K. Bychkov, K. Terebilenko, M. Slobodyanik, Luminescence properties of the new complex La,BiVO$_4$: Mo,Eu compounds as materials for down-shifting of VUV–UV radiation, Radiation Measurements, 2016, v. 90, p. 282-286.

12. S.G. Nedilko, O.V. Chukova, Yu.A. Hizhnyi, S.A. Nedilko, T.A. Voitenko, T. Billot, L. Aigouy, Synthesis and utilization of LaVO$_4$: Eu$^{3+}$ nanoparticles as fluorescent near-field optical sensors, Phys. Status Solidi C, 2015, v. 12(3), p. 282-286.

13. P. Yang, S. Huang, D. Kong, J. Lin, H. Fu, Luminescence functionalization of SBA-15 by YVO$_4$:Eu$^{3+}$ as a novel drug delivery system, Inorg Chem., 2007, v. 46(8), p. 3203-11.

14. K.N. Shinde, R. Singh, S.J. Dhoble, Photoluminescent characteristics of the single-host white-light-emitting Sr3-3x/2(VO4)2:xEu (0<x<0.3) phosphors for LEDs, J. of Luminescence, 2014, v. 146, p. 91-96.

15. Xue Chen, Zhiguo Xia, Luminescence properties of Li$_2$Ca$_2$ScV$_3$O$_{12}$ and Li$_2$Ca$_2$ScV$_3$O$_{12}$:Eu$^{3+}$ synthesized by solid-state reaction method, Optical Materials, 2013, v.35, p. 2736–2739.

16. P. Biswas, V. Kumar, N. Padha, H.C. Swart, Synthesis, structural and luminescence studies of LiSrVO$_4$:Sm$^{3+}$ nanophosphor to fill amber gap in LEDs under n-UV excitation, J Mater Sci: Mater Electron. 2017, v. 28, p. 6159–6168.

17. K. Maheshvaran, K. Linganna, K. Marimuthu, Composition dependent structural and optical properties of Sm$^{3+}$ doped boro-tellurite glasses, J. of Luminescence, 2011, v. 131, p. 2746–2753.

18. O. Chukova, S. Nedilko, S. Zayets, R. Boyko, P. Nagornyi, M. Slobodyanik, Luminescent spectroscopy of sodium titanium orthophosphate crystals doped with samarium and praseodymium ions, Optical Materials, 2008, v. 30, p. 684-686.

19. S. Nedilko, Yu. Hizhnyi, O. Chukova, P. Nagornyi, R. Bojko, V. Boyko, Luminescent monitoring of metal dititanium triphosphates as promising materials for radioactive waste confinement, J. of

Nuclear Materials, 2009, v. 385, p. 479-484.
20. M. Sobszyk, D. Szymański, A study of optical properties of $Sm^{3+}$ ions in $\alpha$-$Na_3Y(VO_4)_2$ single crystals, J. of Luminescence, 2013, v. 142, p. 96-102.
21. O.V. Chukova, S.G. Nedilko. A.A. Slepets, S.A. Nedilko, T.A. Voitenko, Crystal field effect on luminescent characteristics of Europium doped orthovanadate nanoparticles, Proceedings of the 2017 IEEE 7th International Conference on Nanomaterials: Applications and Properties, 2017, v. NAP 2017, p. 81903497-5.
22. V. Chornii, O. Chukova, S.G. Nedilko, S.A. Nedilko, T. Voitenko, Enhancement of emission intensity of $LaVO_4$:$RE^{3+}$ luminescent solar light absorbers, Phys Status Solidi C, 2016, v. 13(1), p. 40–46.
23. O. Chukova, S.A. Nedilko, S.G. Nedilko, V. Sherbatsky, T. Voitenko, Comparable structural and luminescent characterization of the $La_{1-x}Eu_xVO_4$ solid solutions synthesized by solid state and co-precipitation methods, Solid State Phenomena, 2013, v. 200, p. 186-192.
24. T. Higuchi, Y. Hotta, Y. Hikita, S. Maruyama, Y. Hayamizu, H. Akiyama, H. Wadati, D. G. Hawthorn, T. Z. Regier, R. I. R. Blyth, G. A. Sawatzky, H. Y. Hwang, $LaVO_4$:Eu Phosphor Films with Enhanced Eu Solubility, Appl. Phys. Lett., 2011, v. 98, p. 071902-3.
25. Chun‐Jiang Jia Ling‐Dong Sun Zheng‐Guang Yan Yu‐Cheng Pang Shao‐Zhe Lü Chun‐Hua Yan, Monazite and Zircon Type $LaVO_4$:Eu Nanocrystals – Synthesis, Luminescent Properties, and Spectroscopic Identification of the $Eu^{3+}$ Sites, European Journal of Inorganic Chemistry, 2010, v.18, p. 2626-2635.
26. Nian Wang, Wen Chen, Quanfei Zhang, Ying Dai, Synthesis, luminescent, and magnetic properties of $LaVO_4$:Eu nanorods, Materials Letters, 2008, v. 62, p. 109-112.
27. Li-Ping Wang, Li-Miao Chen, Controllable synthesis and luminescent properties of $LaVO_4$: Eu nanocrystals, Materials Characterization, 2012, v. 69, p. 108-114.
28. Yaqiong Zhu, Yonghong Ni, Enhong Sheng, Mixed-solvothermal synthesis and applications in sensing for $Cu^{2+}$ and $Fe^{3+}$ ions of flowerlike $LaVO_4$:$Eu^{3+}$ nanostructures, Materials Research Bulletin, 2016, v. 83, p. 41-47.
29. Jan W. Stouwdam, Mati Raudsepp, Frank C. J. M. van Veggel, Colloidal Nanoparticles of $Ln^{3+}$-Doped $LaVO_4$ : Energy Transfer to Visible- and Near-Infrared-Emitting Lanthanide Ions, Langmuir 2005, v. 21, p. 7003-7008.
30. Reena Okram, Ganngam Phaomei, N. Rajmuhon Singh, Water driven enhanced photoluminescence of Ln (= $Dy^{3+}$, $Sm^{3+}$) doped $LaVO_4$ nanoparticles and effect of $Ba^{2+}$ co-doping, Materials Science and Engineering B, 2013, v. 178, p. 409–416.
31. Junfeng Liu and Yadong Li, Synthesis and Self-Assembly of Luminescent $Ln^{3+}$-Doped $LaVO_4$ Uniform Nanocrystals, Adv. Mater. 2007, 19, 1118–1122.
32. Cailing Chen, Ying Yu, Chunguang Li, Dan Liu, He Huang, Chen Liang, Yue Lou, Yu Han, Zhan

Shi, and Shouhua Feng, Facile Synthesis of Highly Water-Soluble Lanthanide-Doped t-LaVO$_4$ NPs for Antifake Ink and Latent Fingermark Detection, Small 2017, 1702305-8.
33. Fei He, Piaoping Yang, Dong Wang, Na Niu, Shili Gai, Xingbo Li and Milin Zhang, Hydrothermal synthesis, dimension evolution and luminescence properties of tetragonal LaVO$_4$:Ln (Ln = Eu$^{3+}$, Dy$^{3+}$, Sm$^{3+}$) nanocrystals, Dalton Trans., 2011, 40, 11023–11030.
34. O. Chukova, S.G. Nedilko, S.A. Nedilko, T. Voitenko, O. Gomenyuk and V. Sheludko, Study of temperature behavior of luminescence emission of LaVO$_4$ and La$_{1-x}$Eu$_x$VO$_4$ powders, Solid State Phenomena, 2015, v. 230, p. 153-159.
35. O. Chukova, S.G. Nedilko, A.A. Slepets, S.A. Nedilko, T. Voitenko, Synthesis and properties of the La$_{1-x-y}$Eu$_y$Ca$_x$VO$_4$, (0 ≤ x, y ≤ 0.2) compounds. Nanoscale Res. Lett., 2017, v. 12, p. 340-11.
36. wikipedia.org/wiki/Ionic_radius
37. O.V. Chukova, S.A. Nedilko, S.G. Nedilko, A.A. Slepets, T.A. Voitenko, M. Androulidaki, A. Papadopoulos, E.I. Stratakis, Structure, morphology and spectroscopy studies of La$_{1-x}$RE$_x$VO$_4$ nanoparticles synthesized by various methods, in: Nanophotonics, Nanooptics, Nanobiotechnology, and Their Applications, O. Fesenko, L. Yatsenko (eds.), Springer Proceedings in Physics 221, In Press, Accepted Paper.
38. Z.M. Fang, Q. Hong, Z.H. Zhou, S.J. Dai, W.Z. Weng, H.L. Wan, Oxidative dehydrogenation of propane over a series of low-temperature rare earth orthovanadate catalysts prepared by the nitrate method, Catalysis Letters, 1999, v. 61, p. 39–44
39. P. Parhi, V. Manivannan, S. Kohl, P. McCurdy, Synthesis and characterization of M$_3$V$_2$O$_8$ (M = Ca, Sr and Ba) by a solid-state metathesis approach, Bull. Mater. Sci. (India), 2008, v. 31(6), p. 885–890.
40. G. Liu, X. Duan, H. Li, H. Dong, Hydrothermal synthesis, characterization and optical properties of novel fishbone-like LaVO$_4$:Eu$^{3+}$ nanocrystals, Materials Chemistry and Physics, 2009, v. 115, p. 165–171.
41. D. Errandonea, S.N. Achary, J. Pellicer-Porres, A.K. Tyagi, Pressure-induced transformations in PrVO$_4$ and SmVO$_4$ and isolation of high-pressure metastable phases, Inorg. Chem., 2013, v. 52, p. 5464–9.
42. Yongsheng Liu, Renfu Li, Wenqin Luo, Haomiao Zhu, and Xueyuan Chen, Optical Spectroscopy of Sm$^{3+}$ and Dy$^{3+}$ Doped ZnO Nanocrystals, Spectroscopy Letters, 2010, v. 43, p. 343–349.
43. P.S. May, D.H. Metcalf, F.S. Richardson, R.C. Carter, C.E. Miller, R.A. Palmer, Measurement and analysis of excited-state decay kinetics and chiroptical activity in the transitions of Sm$^{3+}$ in trigonal Na$_3$[Sm(C$_4$H$_4$O$_5$)$_3$] * 2NaClO$_4$ * 6H$_2$O, J. of Luminescence, 1992, v. 51, p. 249-268.
44. S.J. Dhoble, S.K. Raut, N.S. Dhoble, Synthesis and Photoluminescence Characteristics of Rare Earth Activated some Silicate Phosphors for LED and Display Devices, Int. J. of Luminescence and Applications 2015, v. 5(2), p. 178-182.

45. J. E. Munoz-Santiuste, V. Lavin, U. R. Rodriguez-Mendoza, Ch. Ferrer-Roca, D. Errandonea, D. Martinez-Garcia, P. Rodriguez-Hernandez, A. Munoz, M. Bettinelli, Experimental and theoretical study on the optical properties of LaVO$_4$ crystals under pressure, Phys. Chem. Chem. Phys., 2018, v. 20, p. 27314.
46. V. Tsiumra, A. Zhyshkovych, T. Malyi, Y. Chornodolskyy, V. Vistovskyy, S. Syrotyuk, Ya Zhydachevskyy, A. Suchocki, A. Voloshinovskii, Localized exciton luminescence in YVO$_4$:Bi$^{3+}$ Optical Materials, 2019, v. 89, p. 480–487.
47. D. Song, C. Guo, T. Li, Luminescence of the self-activated vanadate phosphors Na$_2$LnMg$_2$V$_3$O$_{12}$ (Ln = Y, Gd), Ceramics International, 2015, v. 41, p. 6518–6524.
48. D. Errandonea, A.B. Garg, Recent progress on the characterization of the high-pressure behaviour of AVO$_4$ orthovanadates, Progress in Materials Science, 2018, v. 97, p. 123–169.
49. S.W. Park, H.K. Yang, J.W. Chung, Y. Chen, B.K. Moon, B.C. Cho, J.H. Jeon, J.H. Kim, Photoluminescent properties of LaVO$_4$ Eu$^{3+}$ by structural transformation. Physica B, 2010, v. 405, p. 4040–4044.